\documentstyle[aps,amssymb,12pt]{revtex}

\begin{document}
\title{Aharonov-Casher phase and persistent current\\
in a polyacetylene ring}
\author{Zhijian Li$^1,$J.-Q. Liang$^{1,2}$, F.-C. Pu$^2$}
\address{$^1$Department of Physics and Institute of Theoretical Physics,\\
Shanxi University, Taiyuan, Shanxi 030006, People's Republic of China\\
$^2$Institute of Physics and center for Condensed Matter Physics,\\
Chinese Academy of Sciences, Beijing 100080, People's Republic of China\\
and Department of Physics, Guangzhou Normal College,\\
Guangzhou 510400, People's Republic of China}
\maketitle

\begin{abstract}
We investigate a polyacetylene ring in an axially symmetric, static electric
field with a modified SSH Hamiltonian of a polyacetylene chain. An effective
gauge potential of the single electron Hamiltonian due to spin-field
interaction is obtained and it results in a Fr\"{o}hlich's type of
superconductivity equivalent to the effect of travelling lattice wave. The
total energy as well as the persistent current density are shown to be a
periodic function of the flux of the gauge field embraced by the
polyacetylene ring.
\end{abstract}

Much attention in the study of mesoscopic systems has concentrated upon
topological effects in multiply connected geometries. As is well known, the
Aharonov-Bohm (AB) effect has led to a number of remarkable interference
phenomena in mesoscopic systems, especially in a ring\cite{1}. A typical
example of the phenomena first observed by B\"{u}ttiker {\it et al. }is the
oscillation of the persistent current in mesoscopic rings threaded by a
magnetic flux\cite{1}\cite{2} which leads to a relative phase on the wave
function of a charged particle due to the U(1) gauge theory. Since the
realization of the Berry phase characterized by a geometric meaning, it has
been predicted that analogous interference effects can be induced by the
geometric phases which originate from the interplay between an electron's
spin and orbital degrees of freedom. Loss {\it et al.} studied the
persistent current in a one-dimensional ring in the presence of a static
inhomogeneous magnetic field considering the coupling between spin and
orbital motion by the Zeeman interaction\cite{3}. Using the imaginary time
path-integral method , they showed that the spin wave function accumulates a
Berry phase which leads to a persistent equilibrium current\cite{3}. The
Aharonov-Casher(AC) effect in mesoscopic systems has also been of great
interest, since it specifically includes the spin degree of freedom. Meir 
{\it et al. }showed that spin orbit interaction(SO) in one dimensional rings
results in an effective magnetic flux\cite{4}. Mathur and Stone then pointed
out that observable phenomena induced by SO interactions are essentially the
manifestation of the AC effect and proposed an experiment to observe the AC
oscillation of the conductance in semiconductor samples\cite{5}.
Subsequently Balatsky, Altshuler\cite{6} and Choi\cite{7} studied the
persistent current related to the AC effect. Inspired by these studies for
textured rings, the AC effect has also been analyzed in connection with the
spin geometric phase. Aronov and Lyanda-Geller considered the spin evolution
in conduction rings, and found that SO interaction results in a spin-orbit
Berry phase which plays an interesting role in the transmission probability%
\cite{8}. However, in realistic systems, the interactions between electrons,
lateral dimension, impurities and the coupling of spins complicate the
situation. Cheung {\it et al. }have systematically studied the persistent
currents in ideal, one-dimensional metallic rings, including temperature and
impurity effects\cite{9}.{\it \ }Experimental observations of persistent
currents have been also reported in an ensemble of \symbol{126}10$^7$ Cu
rings\cite{2}, in single gold rings\cite{10} and in a single loop in a GaAs
heterojunction\cite{11}.

It has long been known that the interaction between electrons and the field
of lattice displacements may be responsible for superconductivity, known as
Fr\"{o}hlich superconductivity\cite{12}. The average electron momentum
corresponds to that of a travelling lattice wave which results in the
persistent current in a one-dimensional lattice with periodic boundary
condition i.e. a ring. The periodic lattice displacement gives rise to an
energy gap in the single-electron spectrum with all levels below the gap
occupied and all above it empty. Fr\"{o}hlich suggested that if, when the
electrons was displaced in $k$ space (see Fig.1) so as to give a current
flow, the gaps follow the displaced Fermi surface, superconductive behavior
might result and is called the fluctuation superconductivity or
paraconductivity. The existence of the energy gap eliminates the possibility
of elastic scattering, provided, the velocity of traveling lattice wave is
sufficiently small. Therefore at low temperature an electric current can
exist without resistance. Originally, Fr\"{o}hlich's mechanism of
superconductivity because of restriction to one dimension was regarded as a
mathematical model without real physical significance. However, Bardeen
pointed out that\cite{13} Fr\"{o}hlich's one-dimensional model rather than a
BCS type of pairing was able to account for the origin of superconductivity
in the experiment of Coleman {\it et al}.\cite{14} who observed an
extraordinary increase in conductivity just above the Peierls soft mode
instability in a one-dimensional organic solid. Fr\"{o}hlich's theory
originally was based on a nearly-free electron model and applied only to
T=0K. This concept i.e. the superconductivity in one-dimensional systems
resulted from a coupling between the electrons and a traveling
macroscopically occupied lattice wave was further developed by Allender,
Bray and Bardeen\cite{15}. In a superfluid there must be macroscopic
occupation of a quantum state that picks out a unique reference frame
describing the velocity $v_s$ of the superfluid\cite{15}. Associated with
each value of $v_s$ there is a whole set of elementary excitations of the
system. When excitations come into equilibrium with the rest frame, a
current $j_s(v_s)$ remains. In superfluid helium the macroscopic occupation
is the Bose condensate of the momentum state and it is the common momentum
of pairs that defines $v_s$ in a superconductor based on pairing. In the
Fr\"{o}hlich model $v_s$ is determined by the velocity of the
macroscopically occupied lattice wave which produces energy gaps at
boundaries of the displaced Fermi surface\cite{15}(see Fig.1 (b)). It was
reported in ref.\cite{16} that the macroscopic velocity $v_s$ can be induced
by the AB magnetic flux in a polyacetylene ring, where the traveling lattice
wave is pinned down due to dimerization and the persistent current is solely
due to the AB phase which leads to a collective shift of the electron
momentum equivalent to that of traveling lattice wave. Motivated by the
formal duality between AB and AC effects (but with very different physical
origin), we in this paper investigate the AC phase which,we will see, also
induces the persistent current in a polyacetylene ring.

Polyacetylene is the simplest linear conjugated polymer. The
thermodynamically stable {\it trans} configuration is sketched in Fig.2
illustrating the $\sigma $ bonding and the $\pi $ bonding in the x-y plane.
A tight-binding model is suitable for describing the $\pi $ electrons in a
polyacetylene chain. A full theory has been given by Su, Schrieffer and
Heeger (SSH)\cite{17}. It is certainly interesting to investigate the Fr$%
\stackrel{..}{o}$hlich-like superconductivity related to the AC effect in
the context of the modified SSH model. We consider a one-dimensional
polyacetylene ring of N electrons in which there is one conduction electron (%
$\pi $ electron) per site with spacing $d$. A infinitely long line charge is
assumed to be set along the axis of the ring i.e. the z axis shown in Fig.3.
The AC setup consists of the spin of $\pi $ electron and the electric field
of the line-charge. The Lagrangian for the $\pi $ electron with arbitrary
polarization of spin can be written in nonrelativistic limit as

\begin{equation}
L=\frac m2\stackrel{.}{\vec{x}}^2+\frac \mu c\stackrel{.}{\vec{x}}\cdot (%
\vec{S}\times \vec{E})-\sum\limits_{n=1}^Nv(x-R_n)
\end{equation}
where m and $\vec{x}$ are the mass and position of the electron,
respectively. $\vec{E}=E_R\hat{e}_r$ represents the radial static electric
field of the line-charge with E$_R=\frac{2\rho }R$ where $\rho $ denotes the
charge per unit length on the line and R is the radius of the polyacetylene
ring. $\mu =g\mu _B$ with $\mu _B=$ $\frac{e\hslash }{2mc}$ being the Bohr
magneton. $g$ is the spin g factor which is taken to be 2 here. $\vec{S}$ is
then the dimensionless spin operator. $v(x-R_n)$ denotes the potential
created by the nth ion with coordinate $R_n$. The physical significance of
the second term in the Lagragian eq.(1) is clear, since a moving magnetic
moment $\mu \vec{S}$ is equivalent to an electric dipole moment $\frac \mu c%
\stackrel{.}{\vec{x}}\times \vec{S}.$ Since we consider a one-dimensional
ring, the trivial potential energy of charged particle in the electric field
has no contribution to the following investigation. The Hamiltonian is

\begin{equation}
\text{\ \hspace{0pt}h}=\frac 1{2m}(\vec{P}-\frac \mu c\vec{S}\times \vec{E}%
)^2+\sum_{n=1}^Nv(x-R_n)
\end{equation}
where $\vec{P}$ is momentum, For the $\pi $ electron confining on the ring,
the Hamiltonian eq.(2) is obtained in a cylindrical coordinate as 
\begin{equation}
h=\frac 1{2m}(-i\hbar \frac 1R\frac d{d\varphi }-\frac \mu cE_RS_z)^2+(\frac %
\mu cE_R)^2(\vec{n}\cdot \vec{S})^2+\sum_{n=1}^Nv(x-R_n)
\end{equation}
where

\begin{equation}
\vec{n}=(\sin \varphi ,-\cos \varphi ,0)
\end{equation}
is an unit vector showing in Fig.3. The azimuthal angle $\varphi $ is
related to the coordinate x obviously by $x=R\varphi $. Considering a
tight-binding model with Wannier function $\phi _l$ per site $l$ such that

\begin{equation}
\lbrack \frac{\vec{P}^2}{2m}+\nu (x-R_l)]\phi _l=\varepsilon \phi _l
\end{equation}
we construct the wave function

\begin{equation}
\psi _s(x)=\xi _s\otimes \psi (x)
\end{equation}
which is the product of the spinor

\begin{equation}
\xi _s=\frac 1{\sqrt{2}}(|\vec{n}\rangle +|-\vec{n}\rangle )
\end{equation}
and spacial wave function

\begin{equation}
\psi (x)=\sum_{l=1}^Nc_l\phi _l(x)
\end{equation}
$|\pm \vec{n}\rangle $ denote the spin coherent states defined by ($\vec{S}%
\cdot \vec{n}$)$|\pm \vec{n}\rangle =\pm \frac 12|\pm \vec{n}\rangle $ ,
Taking into account of the unit vector $\vec{n}$ expressed in eq.(4), we
have the spin coherent states written explicitly as

\begin{eqnarray}
|\vec{n}\rangle &=&\frac{\sqrt{2}}2(e^{-\frac 12i\varphi }|+\rangle +e^{%
\frac 12i\varphi }|-\rangle )  \nonumber \\
|-\vec{n}\rangle &=&\frac{\sqrt{2}}2(e^{-\frac 12i\varphi }|+\rangle -e^{%
\frac 12i\varphi }|-\rangle )
\end{eqnarray}
where $|\pm \rangle $ are the usual spin eigenstates of $S_z$, $S_z|\pm
\rangle =\pm \frac 12|\pm \rangle $.

Averaging on the spin state $\xi _s$, we obtain the effective Hamiltonian as

\begin{equation}
h_{eff}=\frac 1{2m}(\vec{P}+\frac ec\vec{A})^2+\sum_{n=1}^Nv(x-R_n)
\end{equation}
where $\vec{A}\equiv \frac c{2e}(\hbar \frac 1R+\frac \mu cE_R)$\^{e}$%
_\varphi $ is formally equivalent to the vector potential of a magnetic flux
with respect to a charged particle with charge e as shown in ref.\cite{16}.
\^{e}$_\varphi $ denotes the unit vector of azimuthal direction. The
Hamiltonian operator eq.(10) is formally the same as that in ref.\cite{16}
where the $\pi $ electron is in the gauge field of a magnetic flux. We
therefore confirm the duality between AB and AC effects by our model study.
Following ref.\cite{16} we assume that $\psi (x)$ is the eigenfunction of $%
h_{eff},$ such that

\begin{equation}
h_{eff}\psi (x)=E(k)\psi (x)
\end{equation}
and satisfies the usual periodic boundary condition

\begin{equation}
\psi (x+Nd)=\psi (x)
\end{equation}
Using an unitary transformation

\begin{eqnarray}
\psi ^{^{\prime }}(x) &=&\exp [i\frac 1{2\hbar }(\hbar \frac 1R+\frac \mu c%
E_R)x]\psi (x)  \nonumber \\
&=&\exp [i\frac{\Phi _{AC}}{Nd\Phi _0}2\pi x]\psi (x)
\end{eqnarray}
the transformed effective Hamiltonian becomes

\begin{equation}
\text{h}_{eff}^{^{\prime }}=\frac{\vec{P}^2}{2m}+\sum\limits_{n=1}^Nv(x-R_n)
\end{equation}
where

\begin{equation}
\Phi _{AC}=\oint \vec{A}\cdot d\vec{l}=\frac c{2e}Nd(\hbar \frac 1R+\frac \mu
cE_R)
\end{equation}
is the effective flux of the gauge potential embraced by the ring which we
may call the AC flux. $\Phi _0=\frac{ch}e$ is the quantum unit of the flux
for a single electron. The stationary schr\"{o}dinger equation is seen to be 
\begin{equation}
\text{h}_{eff}^{^{\prime }}\psi ^{^{\prime }}=E(k)\psi ^{^{\prime }}
\end{equation}
The energy eigenvalue $E(k)$ and the eigenfunction $\psi (x)$ are determined
by both eq.(11) and the usual periodic boundary condition eq.(12).
Correspondingly $\psi ^{^{\prime }}(x)$ possess a nontrivial boundary
condition

\begin{equation}
\psi ^{^{\prime }}(x+Nd)=\exp [i2\pi \frac{\Phi _{AC}}{\Phi _0}]\psi
^{^{\prime }}(x)
\end{equation}
The vector potential is eliminated in the Hamiltonian h$_{eff}^{^{\prime }}$
and the energy spectrum E(k) is then determined by eq.(16) along with the
nontrivial boundary condition of wavefunction eq.(17). The reason why we
make use of the unitary transformation eq.(13) to calculate the energy
spectrum is that we can therefore adopt Dirac's arguments\cite{18} to
demonstrate the periodicity of spectrum with respect to the magnetic flux
following Byers and Yang\cite{19}. The situation is exactly the same as that
in the analysis of the magnetic flux quantization in superconductivity.
Under the unitary transformation the kinetic momentum in the Hamiltonian $%
h_{eff}$ eq.(10)

\[
\vec{P}_{kin}=\vec{P}+\frac ec\vec{A} 
\]
becomes

\[
\vec{P}_{kin}^{^{\prime }}=\vec{P} 
\]
in the Hamiltonian $h_{eff}^{^{\prime }}$ eq.(14) where $\vec{P}$ denotes
the canonical momentum.

Following the argument of Byers and Yang in their famous paper\cite{19}
explaining the magnetic flux quantization in superconductor, we immediately
conclude that the eigenvalue $E(k)$ should be a periodic function of the AC
flux $\Phi _{AC}$ and an even function of $\Phi _{AC}$. When $\frac{\Phi
_{AC}}{\Phi _0}$ equals an integer, the boundary condition eq.(17) coincides
with the usual one i.e. eq.(12), and then the charged line would not result
in any observable effect similar to the AB case shown in ref.\cite{16} where
the magnetic flux becomes a Dirac string. Using eq.(15) it is easy to find
that the charge per unit length for the line-charge resulting in no
observable effect equals to

\[
\rho =(n+\frac 12)\frac{mc^2}e
\]
where $n$ is an integer. The AC partner of the quantum unit of magnetic flux 
$\Phi _0$ in ref.\cite{16} i.e. in the AB case is seen to be $\rho _0=\frac{%
mc^2}e$ .

To see how the eigenvalue E(k) depends on the AC flux, we evaluate the
single-electron energy band with a small variation of AC flux from the
integral units of $\Phi _0$. The second-quantization Hamiltonian of the
electron can be written as

\begin{eqnarray}
H &=&\int \psi ^{+}(x)h_{eff}\psi (x)dx=\int \psi ^{^{\prime
}+}(x)h_{eff}^{^{\prime }}\psi ^{^{\prime }}(x)dx  \nonumber \\
&=&-\sum_l[t_0-\alpha (u_{l+1}-u_l)](a_{l+1}^{+}a_l+a_l^{+}a_{l+1}) 
\nonumber \\
&&+iMv_f\sum_l(a_{l+1}^{+}a_l-a_l^{+}a_{l+1})  \nonumber \\
&&+\frac 12mv_f^2\sum_la_l^{+}a_l
\end{eqnarray}

The first sum is the usual result in the tight-binding approximation and is
seen to be SSH Hamiltonian for electrons. $t_0$ denotes the hopping integral
in the ion equilibrium position, namely the undimerized chain, while $\alpha 
$ is the electron-phonon coupling constant. The second and the third sum
come from the contribution of the AC flux, where $M\equiv i\int \varphi
_{l-1}^{*}\vec{P}\varphi _ldx$ is the dipole matrix element which is assumed
to be the same for all $l$ and $v_f=\frac{2\hbar \pi }{mNd}\frac{\Phi _{AC}}{%
\Phi _0}$ is regarded as a macroscopic velocity of electrons induced by AC
flux i.e.

\[
v_s=v_f 
\]

For the perfectly dimerized chain, the displacement configuration of the
lattice may be written as $u_n=(-1)^nu$. Following SSH we separate the $a_n$
into those of odd ($a_{n_o}$) and even ($a_{n_e}$) indices and introduce the
Fourier transforms

\begin{eqnarray}
a_{n_o} &=&\frac 1{\sqrt{N}}\sum\limits_k(c_k^v+c_k^c)e^{-i2\pi n_odk} 
\nonumber \\
a_{n_e} &=&\frac 1{\sqrt{N}}\sum\limits_k(c_k^v-c_k^c)e^{-i2\pi n_edk}
\end{eqnarray}
where $\mid k\mid \leq \frac 1{4d}$, the superscripts $v$ and $c$ denote the
valence and conduction bands respectively. The Hamiltonian (10) can be
expressed in the $k$ representation.

\begin{eqnarray}
H &=&\sum\limits_kE_0(k)\left( c_k^{v+}c_k^v-c_k^{c+}c_k^c\right)
+i\sum\limits_k\Delta (k)\left( c_k^{v+}c_k^c-c_k^{c+}c_k^v\right)  \nonumber
\\
&&+2v_fM\sum_k\sin (2\pi kd)\left( c_k^{v+}c_k^c-c_k^{c+}c_k^v\right) 
\nonumber \\
&&+\frac 12mv_f^2\sum_k\left( c_k^{v+}c_k^v+c_k^{c+}c_k^c\right)
\end{eqnarray}
where 
\begin{eqnarray}
E_0 &=&-2t_0\cos (2\pi kd)  \nonumber \\
\Delta (k) &=&4\alpha u\sin (2\pi kd)
\end{eqnarray}

In order to diagonalize the Hamiltonian (20), we make the following
Bogoliubov transformation,

\begin{eqnarray}
a_k^v &=&-i\alpha _kc_k^v+\beta _kc_k^c  \nonumber \\
a_k^c &=&\alpha _k^{*}c_k^v+i\beta _kc_k^c
\end{eqnarray}
with $\mid \alpha _k\mid ^2+$ $\mid \beta _k\mid ^2=1$. The diagonalized
form is

\begin{equation}
H=-\sum\limits_kE(k)\left( a_k^{v+}a_k^v-a_k^{c+}a_k^c\right) +\frac 12%
mv_f^2\sum\limits_k\left( a_k^{v+}a_k^v+a_k^{c+}a_k^c\right)
\end{equation}
where 
\begin{eqnarray}
E(k) &=&\sqrt{E_0^{^{\prime }}(k)^2+\Delta (k)^2}  \nonumber \\
E_0^{^{\prime }}(k) &=&E_0(k)+2v_fM\sin 2\pi kd
\end{eqnarray}
A real solution for the transformation coefficient in eq.(22) is seen to be

\begin{eqnarray}
\alpha _k &=&\frac 1{\sqrt{2}}\sqrt{1+\frac{E_0^{^{\prime }}(k)}{E(k)}} 
\nonumber \\
\beta _k &=&-\frac 1{\sqrt{2}}\frac k{\mid k\mid }\sqrt{1-\frac{%
E_0^{^{\prime }}(k)}{E(k)}}
\end{eqnarray}

At zero temperature the valence band is occupied while the conduction band
is empty. The total energy is obtained as

\begin{equation}
U=-2\sum\limits_{k=-1/4d}^{1/4d}\sqrt{E_0^{^{\prime }}(k)^2+\Delta ^2(k)}+%
\frac 12Nmv_f^2
\end{equation}
Under the condition $\frac{v_fM}{t_0}\ll \frac{2\alpha u}{t_0}\ll 1$, the
first term can be approximated up to quadratic dependence on the small
variation of AC flux away from an integer value of $\Phi _0$. Since the flux 
$\Phi _{AC}$ whenever becoming an integral multiple of $\Phi _0$ will not
affect the energy eigenvalue $E(k)$ i.e. $v_f$ vanishes, the total energy
would be a minimum seeing from eq.(26). The total energy of the occupied
valence band has infinite number of minima wherever the magnetic flux is an
integral multiple of $\Phi _0$. The AC flux dependence of the energy is
shown in Fig.4.

To obtain the macroscopic momentum shift, namely the displacement of
electrons in $k$ space,we first have to find the electron wave function for
the single-electron Hamiltonian $h_{eff}^{^{\prime }}$. Such a wave function
which leads to the valence band $E_v=-E(k)$ is

\begin{equation}
\psi ^{^{\prime }}(x)=\frac{\exp (i\frac{mv_f}\hbar x)}{\sqrt{N}}\left(
\alpha _k\sum\limits_ne^{i2\pi knd}\phi _n+i\beta _k\sum\limits_ne^{i2\pi
knd}(-1)^n\phi _n\right)
\end{equation}
The expectation value of kinetic momentum $\langle \vec{P}_{kin}\rangle $
can be calculated with the above wave function,

\begin{eqnarray}
\left\langle \vec{P}_{kin}\right\rangle &=&\langle \psi |\vec{P}+\frac ec%
\vec{A}|\psi \rangle  \nonumber \\
&=&\langle \psi ^{^{\prime }}|\vec{P}|\psi ^{^{\prime }}\rangle  \nonumber \\
&=&mv_f+\frac{2ME_0^{^{\prime }}(k)}{E(k)}\sin (2\pi kd)
\end{eqnarray}
where the contribution from the second term is negligible when summed over $%
k $. The persistent current density due to a small deviation of AC flux from
an integral number of $\Phi _0$ is seen to be

\begin{equation}
j=nev_f
\end{equation}
where n=$\frac 1d$ is the electron density . It is seen that the current
density depends on AC flux periodically and discontinuous at half-integral
values of $\Phi _0,$ which is shown in Fig.5.

In summary, we study a polyacetylene ring in existence of the radial
electric field induced by an infinitely long charge line in the
one-dimensional tight-binding model with the electron-phonon interaction. An
effective gauge potential of $\pi $ electron due to spin-field interaction
leads to Fr\"{o}hhich's type of superconductivity equivalent to the effect
of travelling lattice wave. The flux of the gauge potential which results in
a collective momentum shift of electron leads to a persistent current, which
oscillates with respect to the AC flux and may be observed experimentally
with a mesoscopic polyacetylene tube.

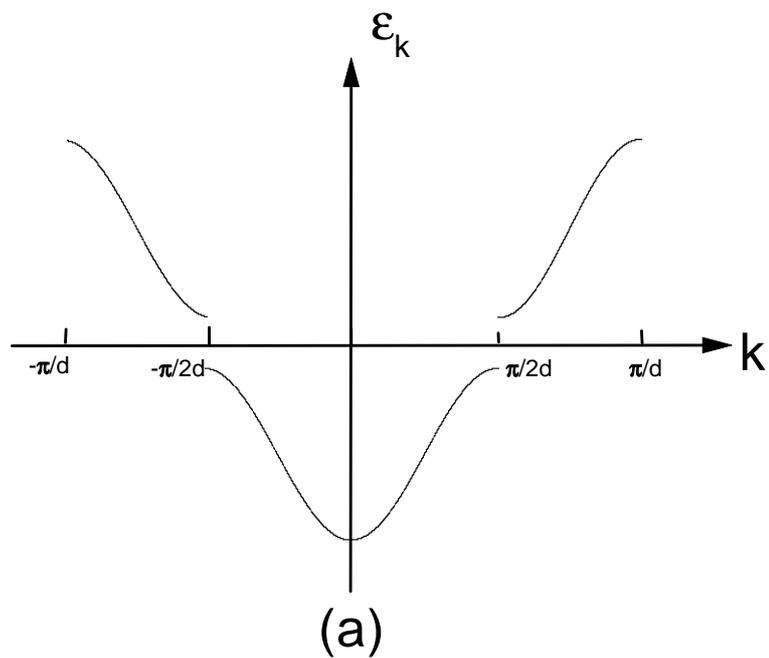
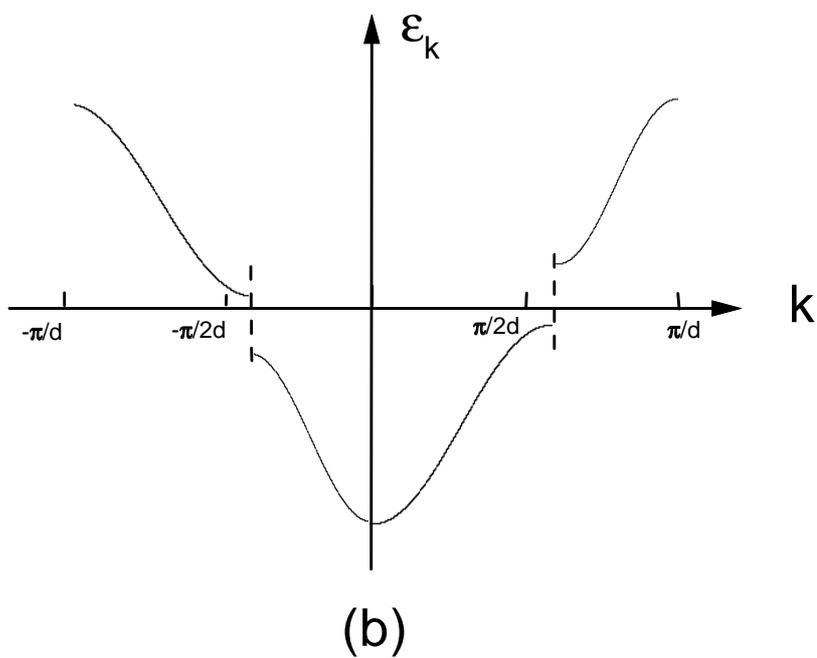
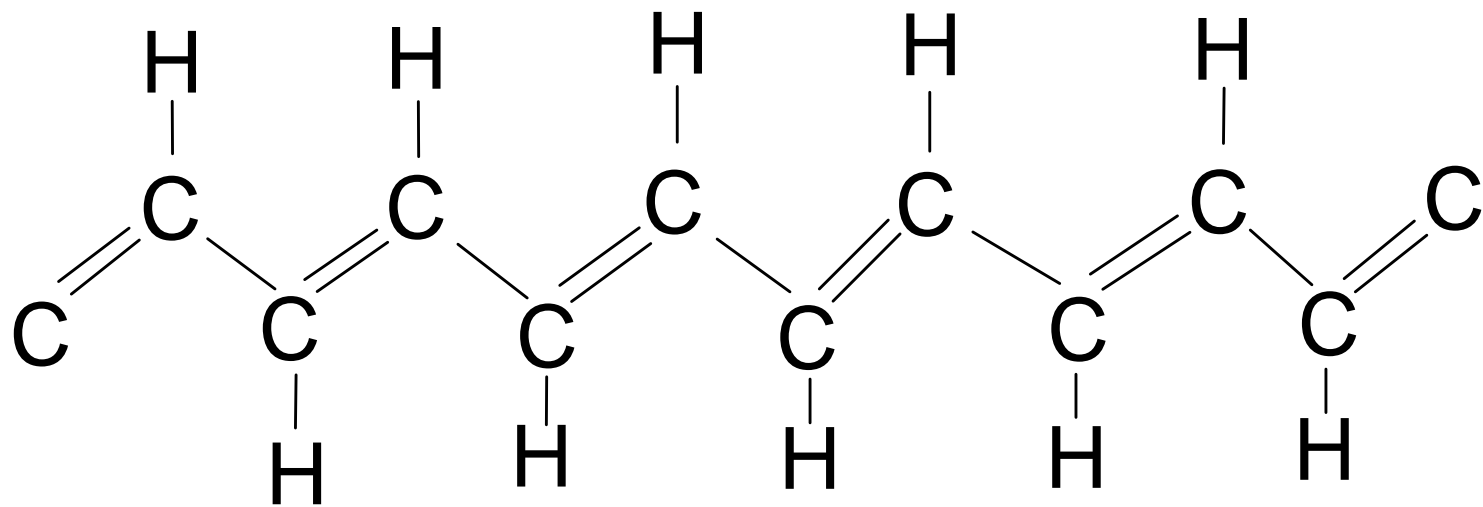
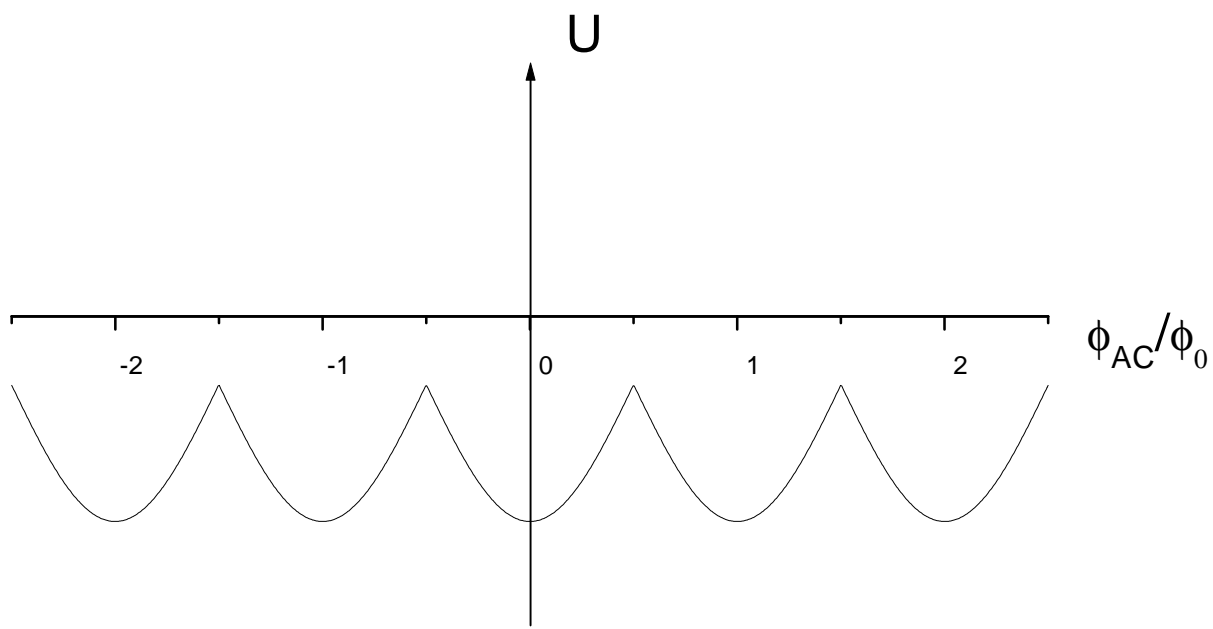
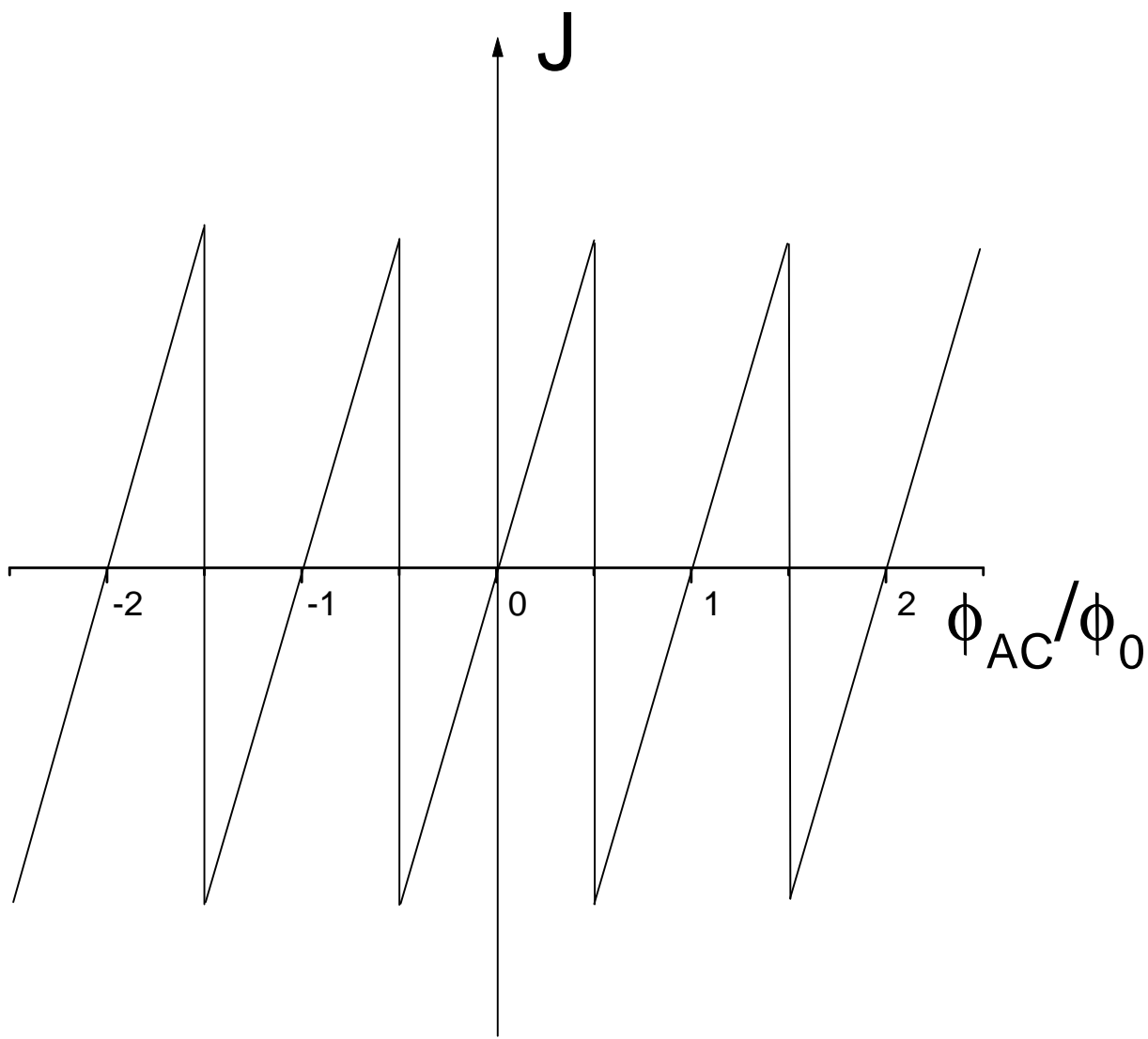
\end{document}